\documentclass[12pt,preprint]{aastex}

\def\apj{{ApJ}}
\def\apjl{{ApJL}}

\usepackage{epsfig}
\usepackage{url}
\usepackage{threeparttable}

\begin{document}
\newcommand{\frb}{FRB~131104}
\newcommand{\gr}{$\gamma$-ray }
\newcommand{\grs}{$\gamma$-rays}
\newcommand{\gcn}{GCN Circ.}


\shorttitle{GeV afterglow from FRBs}
\title{Search for GeV counterparts to fast radio bursts with \textsl{Fermi}}
\shortauthors{}
\author{
Shao-Qiang Xi\altaffilmark{1,3}, Pak-Hin Thomas
Tam\altaffilmark{2}, Fang-Kun Peng\altaffilmark{1,3} and Xiang-Yu
Wang\altaffilmark{1,3}}

\affil{$^1$ School of Astronomy and Space Science, Nanjing
University, China; xywang@nju.edu.cn\\
$^2$ School of Physics and Astronomy, Sun Yat-sen University,
Zhuhai 519082, China; tanbxuan@mail.sysu.edu.cn\\
$^3$Key laboratory of Modern Astronomy and Astrophysics, Nanjing
University, Ministry of Education, Nanjing 210093, China}

\begin{abstract}
The non-repeating  fast radio bursts (FRBs) could arise from catastrophic stellar explosions or magnetar giant flares, so relativistic blast waves might be produced in these events. Motivated by this,  we here search for GeV counterparts to all non-repeating FRBs with \textsl{Fermi} Large Area Telescope (LAT), including FRB 131104 that is claimed to be possibly associated with a \gr transient candidate detected by \textsl{Swift} Burst Alert Telescope (BAT).   FRB 131104 enters the
field of view (FoV) of LAT $\sim 5000 \ {\rm s}$ after the burst time, so we are only able to
search for the GeV
afterglow emission during this period, but no significant GeV emission is detected.
we also perform a search for GeV emissions from other FRBs, but
no significant GeV emissions are detected either.
Upper limit fluences in  the range of (4.7--29.2)$\times 10^{-7}\ {\rm erg \ cm^{-2}}$ are obtained, and then the upper limits of the isotropic blast wave kinetic energy of about (1--200)$\times 10^{53} \ {\rm erg}$ are inferred under certain assumptions. Although the current limits on the isotropic blast wave energy are not sufficiently stringent to rule out the connection between FRBs and GRB-like transients, future  more sensitive observations with \textsl{Fermi} or Imaging Atmospheric  Cherenkov Telescopes might be able to constrain the connection.

\end{abstract}

\keywords{gamma rays: bursts ---
                gamma rays: observations}

\section{Introduction}
 Fast radio bursts (FRBs) are  intense bursts of radio emission that have durations of milliseconds and have large dispersion measures.
They were first discovered in the
archival data from the Parkes telescope, as reported
in~\citet{2007Sci...318..777L} and \citet{2013Sci...341...53T},
and subsequently by other radio telescopes as
well~\citep{2015MNRAS.447..246P}. The origin of these radio bursts
is still an enigma, even though more than a dozen FRBs have been
found\footnote{\url{http://www.astronomy.swin.edu.au/pulsar/frbcat/}}
\citep{2016PASA...33...45P}.

Their large dispersion measures (DMs) exceed predictions for the Galaxy, suggesting that the FRBs are extragalactic and possibly cosmological in origin \citep{2013Sci...341...53T,2016Sci...354.1249R}. The DM contributed by the host galaxy, the possible intervening galaxy and the plasma around the FRB sources is unknown, so the distances of the FRB sources are uncertain.
Recently, owing to the multi-wavelength follow-up observations of the host galaxy of the repeating source FRB 121102 \citep{2017Natur.541...58C,2017ApJ...834L...8M,2017ApJ...834L...7T}, the distance scale of this FRB is finally settled to be cosmological ($\sim$ Gpc).
At these distances, the isotropic-equivalent energy output of a typical FRB
is about 10$^{38}$ to 10$^{42}$ erg. Such energetics, together
with the millisecond duration, suggest that FRBs are likely
related to compact objects, including non-catastrophic, sometimes
repeatable events \citep[e.g.,][]{2010vaoa.conf..129P,2014MNRAS.439L..46L,
2014ApJ...797...70K,2015ApJ...809...24G,
2016ApJ...822L...7W,2016MNRAS.458L..19C,2016ApJ...826..226K,
2016ApJ...829...27D}, and catastrophic events such as compact star
mergers \citep[e.g.,][]{2013PASJ...65L..12T,2013ApJ...776L..39K,2014ApJ...780L..21Z,
2015ApJ...814L..20M,2016ApJ...827L..31Z}.

Such merger systems are expected to produce relativistic blast waves, and in turn produce multi-wavelength afterglows \citep{2014PASJ...66L...9N,2014ApJ...792L..21Y,2016MNRAS.461.1498M}.  The giant flares of soft \gr repeaters (SGRs)
also produce blast waves and thus long-term radio emission, as observed in SGR 1806-20 \citep{1999Natur.398..127F,2005Natur.434.1104G,2005Natur.434.1112C,2005ApJ...623L..29W}.
There have been  efforts to search for
electromagnetic counterparts to FRBs. For example, FRB 150418 was widely searched at energy range from radio to very high-energy \gr (TeV), but it yielded null result~\citep{2016ApJ...821L..22W,2016ApJ...824L...3A,
2017A&A...597A.115H}.
 It was recently reported that a \gr transient candidate
was in coincident with the FRB 131104, with an
association significance of about 3.2$\sigma$ \citep{2016ApJ...832L...1D}. The transient is located near the edge of the BAT'
field of view, leading to a low-significance signal in spite of a
relatively bright fluence of $S_\gamma\simeq 4\times 10^{-6} \ {\rm
erg \ cm^{-2}}$, which is comparable to that of cosmological
\gr bursts (GRBs). However, \citet{2017ApJ...837L..22S} argue that the association between the \gr transient and the FRB is not compelling due to the non-detection of radio afterglow emission
at the original location region of the \textsl{Swift}/BAT transient and  the discovery of a radio AGN spatially and temporally coincident with FRB 131104. Interestingly, Bannister et al. (2012)
searched the radio pulse emission from nine gamma-ray bursts and detected single dispersed pulses following two GRBs at significance $>6 \sigma$.
A simple population argument supports a GRB origin with confidence of about $2\%$ and they cannot rule out radio frequency interference  as the origin of these pulses.

If the association between FRB 131104 and the \textsl{Swift}/BAT transient is
real, this would open up the possibility that FRBs may be
accompanied by relativistic shock phenomena similar to GRB
afterglows. Multi-wavelength afterglow emission is then expected, but searches for the radio, optical and X-ray afterglows have been unsuccessful.
The intensity of the radio to X-ray afterglow
emission depends on the density of the circumburst medium. Using
the non-detection, several groups have constrained the density of
the medium to be very tenuous, i.e. $n \la 10^{-3} \ {\rm cm^{-3}}$
\citep{2016arXiv161109517D,2017ApJ...836L...6M,
2017ApJ...835L..21G}. On the other hand, we note that GeV
afterglow emission is independent of the circumburst density, so
search for GeV afterglow emission using \textsl{Fermi}-LAT would
be useful to test the relativistic shock scenario, independent of
the circumburst environment. Furthermore, several GRBs with a
fluence comparable to the \textsl{Swift}/BAT transient possibly associated with
FRB 131104 have been found to emit prompt or/and afterglow GeV emission. So we also
attempt to search for the GeV prompt or/and afterglow emission associated with
FRBs, assuming that FRBs are associated with some types of GRBs.

\citet{2016MNRAS.460.2875Y} has focused
on a blind search of millisecond \gr flashes over the whole set of
8-year LAT data, without regarding to the known FRB triggers,
implicitly assuming any GeV counterparts to FRBs are of similar
millisecond duration. In contrast to \citet{2016MNRAS.460.2875Y}, our current work is to search for any GeV emission counterparts of the reported non-repeating FRBs in the literatures, utilizing data taken from the LAT observations onboard the \textsl{Fermi} satellite.
In Section 2 we describe the \textsl{Fermi}-LAT data analysis. In Section 3 we discuss our data analysis results. The simple summary is given in Section 4.
\section{\textsl{Fermi}-LAT data analysis}

\textsl{Fermi}-LAT, the primary instrument on board the \textsl{Fermi} Gamma-ray Space Telescope, is an imaging, wide field-of-view (2.4 sr), and high-energy (20 MeV -- $\sim$ 300~GeV) detector. For more details about the LAT, the reader is referred to \cite{2009ApJ...697.1071A}. The effective area and energy range increase in the latest Pass 8 data\footnote{\url{http://www.slac.stanford.edu/exp/glast/groups/canda/lat\_Performance.htm}}. Utilizing the newly released \textsl{Fermi}-LAT
 Pass  8 data and \textsl{Fermi}
Science Tools (v10r0p5) with the P8R2\_SOURCE\_V6 instrument response functions,
we search for GeV prompt or/and afterglow emission of the potential \gr transient associated with non-repeating FRBs. We consider all SOURCE class events (i.e., evclass=128 and evtype=3) in the energies between 100 MeV and 100 GeV within a 10 degree region of interest (ROI) centered on the each FRB position. Further more, we use a zenith angle cut of $Z_{\rm max}<90^{\circ}$ to greatly reduce contamination by the Earth limb emission and apply the recommended data-quality cuts of (DATA\_QUAL $> 0$) \&\& (LAT\_CONFIG ==1). The unbinned likelihood analysis is
performed with a source model including a point source with a
power-law spectrum ($dN/dE=A\times (E/E_0)^{-\Gamma}$) on the each FRB position, the two diffuse emission components, i.e., Galactic diffuse emission model (gll\_iem\_v6.fits) and the isotropic diffuse emission model (iso\_P8R2\_SOURCE\_V6\_v06.txt), and
the 3FGL sources \citep{2015ApJS..218...23A} within $15^{\circ}$ from the ROI center. The time interval for \textsl{Fermi}-LAT analysis is selected when the FRB region enters the field of view of \textsl{Fermi}-LAT after FRB detection time, and the start and stop times are given by the angle $\Theta < 70^{\circ}$, which is the angular distance between the FRB position and \textsl{Fermi}-LAT boresight. The analysis time interval for each FRB is listed in Table \ref{FRBdata1}. For the analysis of short time period of a
few thousand seconds, the normalization factors of all 3FGL
sources is left to vary and the spectral indices are frozen to their
catalog values to solve convergence problems. No significant high-energy \gr emission at the FRB positions are found, and thus we get the upper limit fluences at $95\%$ confidence level with fixed spectral indices of
$\Gamma = 2.2$. We have checked our analysis results with the transient data (using the corresponding response function IRFs=P8R2\_TRANSIENT020\_V6) or different spectral index, and no significant difference is found.

The fluence of the possible \gr transient associated with FRB 131104 reached $\sim 10^{-6} \ {\rm erg \ cm^{-2}}$, which is comparable to the the prompt fluence of typical GRBs. For comparison, we try to  check whether the \textsl{Swift} GRBs, which have the similar order of magnitude of the prompt fluence level, show any GeV emission. Using the tool \emph{gtburst} provided in the software package \textsl{Fermi} Science Tools v10r0p5\footnote{\url{https://fermi.gsfc.nasa.gov/ssc/data/analysis/scitools/gtburst.html}}, we carry out a standard maximum likelihood analysis of \textsl{Fermi}-LAT GRB data in the time interval when the main GeV emission radiates. We select events of TRANSIENT class for short time interval($< 100 \ {\rm s}$) and SOURCE class for longer time interval, with the corresponding P8R2\_TRANSIENT020\_V6 and P8R2\_SOURCE\_V6 instrument response functions respectively. The size of ROI of each GRB is $12^{\circ}$. We excluded the Earth Limb emission with the zenith angles cut of $Z_{\rm max}< 90^{\circ}$.

\begin{figure}
\centering
\includegraphics[angle=0,scale=0.5]{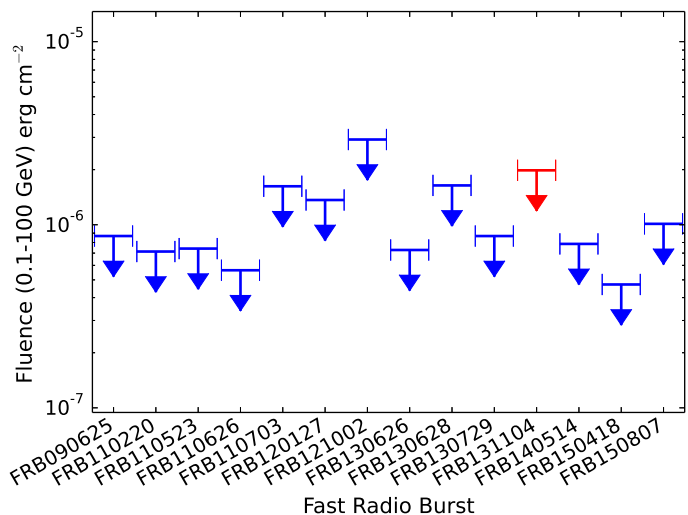}
\caption{Upper limit fluence of each FRB in 0.1--100~GeV. The red one represents FRB 131104, which is possibly associated with the \textsl{Swift}/BAT transient. }
\label{FRBfig1}
\end{figure}

\begin{figure}
\centering
\includegraphics[angle=0,scale=0.5]{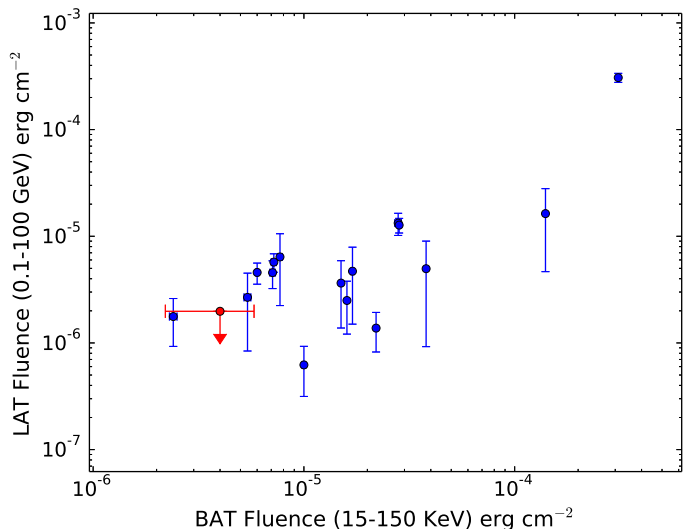}
\caption{\textsl{Fermi}-LAT fluences vs. \textsl{Swift}/BAT fluences for GRBs detected by both instruments. For comparison, the red data is for FRB 131104 which is possibly associated with the \textsl{Swift}/BAT transient. }
\label{FRBfig2}
\end{figure}

The upper limit GeV fluences of point source centered on each FRBs position are presented in Table \ref{FRBdata1} and Figure \ref{FRBfig1}. One can see that no FRB shows any significant  GeV emission, which could be used to constrain the energy of relativistic blast waves  in some FRB models based on certain assumptions (see Section 3). 16 \textsl{Swift}/BAT GRBs are observed simultaneously
by \textsl{Fermi}-LAT. The results of \textsl{Swift} GRB sample are listed in Table \ref{FRBdata2}. Several of them have the BAT fluence levels comparable to the possible \gr transient associated with FRB 131104, but they have GeV emission (also see Figure \ref{FRBfig2}).

%
\section{Discussions}
The association between FRB 131104 and the \textsl{Swift}/BAT transient is still controversial \citep{2016ApJ...832L...1D,2017ApJ...837L..22S}. The radio-continuum imaging observations of the localization region of the FRB do not find any radio afterglow of this transient \citep{2017ApJ...837L..22S}, which puts the cosmic fireball model into question. Nevertheless, since the radio afterglow flux is sensitive to the circumburst medium density, the non-detection of radio emission could be  due to a low circumburst density. Given these uncertainties, we tentatively assume that the FRB 131104 or some other FRBs  are associated with  gamma-ray transients.
If the association between FRB 131104 and the \textsl{Swift}/BAT transient is
true, the energy of this event would be comparable to cosmological GRBs.
Some GRBs  with a comparable fluence, such
as GRB~110731A, GRB~120729A, and GRB~150512A (see Table \ref{FRBdata2}),  have
been detected by \textsl{Fermi}-LAT. For comparison, we show, in Figure \ref{FRBfig2},
the relation between the LAT fluences and BAT fluences for those
GRBs that were detected by both \textsl{Fermi}-LAT and \textsl{Swift}/BAT. As the BAT
transient associated with FRB 131104 has a relatively low fluence
in BAT energy window, it is not surprising that this event was not
detected by \textsl{Fermi}-LAT, especially considering that this event
enters the FoV of LAT about 5000 s after the radio burst.

 Expanding blast waves driven by cosmic fireballs, such as GRBs, may
accelerate relativistic electrons through Fermi acceleration, which in turn produce long-term  synchrotron emission in the magnetic field,
the so-called "afterglows" \citep{1997ApJ...476..232M,1998ApJ...497L..17S}. Assuming a power-law electron distribution with an index of $p$, the  flux density of the afterglow synchrotron emission
at GeV energies is given by \citep{2009MNRAS.400L..75K}
\begin{eqnarray}\label{afterglowEq}
F_\nu&=& 0.2 \ {\rm mJy} E_{55}^{(p+2)/4}\epsilon_{e}^{p-1} \epsilon_{B,-2}^{(p-2)/4}t_{1}^{-(3p-2)/4} \nonumber \\
&&\times\nu_{8}^{-p/2}(1+z)^{(p+2)/4}d_{L28}^{-2},
\end{eqnarray}
where $E$ is the isotropic kinetic energy of the blast wave, $\epsilon_{e}$ and $\epsilon_{B}$ are, respectively, the
fractions of energy of the shocked gas in electrons and magnetic
fields, $t_{1}\equiv t/10 \ {\rm s}$ is the time since the beginning of the explosion in
the observer frame, $\nu_{8}$ is photon energy in units of 100
MeV, $z$ is the
redshift, and $d_{L28}\equiv d_{L}/10^{28} \ {\rm cm}$ is the luminosity
distance to the burst. Using this
formula, we can obtain the upper limit of the  blast wave kinetic
energy of each FRB, assuming that  FRBs are at distances
corresponding to the measured DM values, and assuming typical
reference values for the shock microphysical parameters (i.e.,
taking $p=2.4$, $\epsilon_{e}=0.1$, and $\epsilon_{B}=0.01$).

With a limit average flux at GeV energies at 5417--11142 s after
the radio burst of FRB 131104, we obtain an upper limit of the
isotropic kinetic energy of the possible blast wave, i.e. $E_k\la
10^{55} \ {\rm erg}$, assuming a luminosity distance corresponding to non-Galactic
$ {\rm DM} \simeq 707.9 \ {\rm pc \ cm^{-3}}$ that totally arises from intergalactic medium. For all our FRBs sample, we find that the upper limits of blast wave energy are in the range of (0.4--190.8)$\times 10^{53} \ {\rm erg}$, which are obtained with upper limit GeV  fluences in the range of (4.7--29.2)$\times 10^{-7} \ {\rm erg \ cm^{-2}}$.
Note that the contribution to the DM of FRBs  by their local environment and host galaxy may decrease the value of the luminosity distance. Then we find that the energy of the blast wave may decrease correspondingly, according to Eq.(\ref{afterglowEq}). For example, if the real luminosity distance  is a half  of that inferred from the non-Galactic DM\footnote{\cite{2017ApJ...834L...7T} reported that the host galaxy of FRB 121102 is at a redshift of $z = 0.19273(8)$. The real luminosity distance is $972 \ {\rm Mpc}$, which is a fraction of $58\%$ of the inferred luminosity distance($ \simeq 1.66 \ {\rm Gpc}$) from the non-Galactic DM.}, the upper limit value of the blast wave energy will decrease by about $70\%$. Therefore, although current limits obtained for the blast wave energy are not sufficiently stringent to rule out the GRB-FRB connection, future more sensitive observations with {\em Fermi} or Imaging Atmospheric  Cherenkov Telescopes at TeV energy could be useful to constrain the connection.

\section{Summary}
It was recently reported that a transient \gr counterpart to
the FRB 131104 was discovered in \textsl{Swift}
satellite data, which, if true, would increase the energy budget
of FRBs to a level comparable to that of cosmological \gr
bursts. The relativistic blast wave driven by such amount
of energy may produce multi-wavelength afterglow emission, but
searches for radio, optical and X-ray afterglows from FRB 131104
have so far not resulted in positive detection. It has been argued that the non-detection of
radio to X-ray afterglow follows from the fact that the event
occurs in a low-density environment\citep[e.g.,][]{2016arXiv161109517D,2017ApJ...836L...6M,
2017ApJ...835L..21G}. In contrast, high-energy
\gr afterglow flux is not sensitive to the
circumburst density, so we  searched for possible GeV afterglows
of FRBs with \textsl{Fermi} Large Area Telescopes.  While
several FRBs were within the field of view of LAT at the burst
time, FRB 131104 entered into the LAT FoV only about 5000 seconds
after the radio burst. No GeV emission is found during this period
for FRB 131104. For those other FRBs in our sample, we also search for possible GeV emission at the time
immediately after the radio burst, but no GeV emission is found
either. With the upper limit fluences at GeV energies, we are able
to obtain upper limits on the kinetic energy of relativistic
blast waves that are possibly associated with these FRBs. The current limits are
not so stringent enough that can be used to constrain the connection between FRBs and GRB-like transients. Nevertheless, future more sensitive observations with \textsl{Fermi} or Imaging Atmospheric  Cherenkov Telescopes, such as CTA, might be able to constrain the connection.

\section*{Acknowledgments}

We thank Zi-Gao Dai for useful discussions. This research made use of data supplied by the High Energy Astrophysics Science Archive Research Center (HEASARC) at NASA's Goddard Space Flight Center, and the UK \textsl{Swift} Science Data Centre at the University of Leicester.
This work is supported by the 973 program under grant 2014CB845800 and the National Natural Science Foundation of China (NSFC) under grants 11625312 and 11273016. PHT is supported by the NSFC grants 11633007 and 11661161010.

\clearpage

\begin{table}
\centering
\begin{threeparttable}
\caption{ Upper limits of the \gr fluences of FRBs observed by \textsl{Fermi}-LAT.}
\begin{tabular}{lcccccccc}
\hline
\hline
Name & (l,b)\tablenotemark{*} & $T+T_{0}$\tablenotemark{**}  &  Fluence limit              & $E_{\rm K}$ limit & ${\rm DM_{non-Galaxy}}$\tablenotemark{*}  & z\tablenotemark{*}\\
     &  ($^{\circ}$)   &(s) &  ($\times 10^{-7}$ erg cm$^{-2}$) & ($\times 10^{53}$ erg) & $\rm pc \ cm^{-3}$ & \\
\hline
FRB131104 & (260.5,-21.9) & [5417,11142] & 19.8 & 190.8 & 707.9 & 0.59 \\
\hline
FRB090625 & (226.4,-60.0) & [0,595] & 8.7 & 7.19 & 867.8 & 0.72 \\
FRB110703 & (80.9,-59.0) & [0,1660] & 16.2 & 16.7 & 1071.2 & 0.89 \\
FRB121002 & (308.2,-26.2) & [0,5119] & 29.2 & 51.6 & 1554.9 & 1.3 \\
FRB130628 & (225.9,30.6) & [0,305] & 16.4 & 3.7 & 417.3 & 0.35 \\
FRB150418 & (232.6,-3.2) & [0,1193] & 4.7 & 1.9 & 587.7 & 0.49 \\
FRB150807 & (336.7,-54.4) & [0,4946] & 10.1 & 0.4 & 196.5 & 0.16 \\
\hline
FRB110220 & (50.8,-54.7) & [2488,4978] & 7.1 & 99.6 & 909.6 & 0.76 \\
FRB110523 & (56.1,-37.8) & [1487,2687] & 7.4 & 44.2 & 579.7 & 0.48 \\
FRB110626 & (355.8,-41.7) & [4622,6152] & 5.6 & 115.9 & 675.5 & 0.56 \\
FRB120127 & (49.2,-66.2) & [1514,3284] & 13.6 & 50.6 & 521.4 & 0.43 \\
FRB130626 & (7.4,27.4) & [2085,4305] & 7.3 & 88.8 & 885.5 & 0.74 \\
FRB130729 & (324.7,54.7) & [302,2842] & 8.7 & 23.5 & 830 & 0.69 \\
FRB140514 & (50.8,-54.6) & [1688,3488] & 7.9 & 34.7 & 527.8 & 0.44 \\
\hline

\end{tabular}

\begin{tablenotes}
\item\textbf{Notes.} The first column represents the FRB names. The second column represents the position in the Galactic coordinate system. The third column represents the analysis time intervals relative to the FRB detection times.
The fourth column represents the upper limit fluences in 0.1--100~GeV. The fifth column represents the estimated upper limits of the isotropic blast wave energy. The last two columns represent the FRB non-Galactic DM and the redshift derived from this non-Galactic DM value respectively.
\item[*] \url{http://www.astronomy.swin.edu.au/pulsar/frbcat/}
\item[**] The analysis time interval is selected requiring that the angular distance between the FRB position and the \textsl{Fermi}-LAT boresight is to be $\le 70 ^{\circ}$. $T_{0}$ is referred to each FRB detection time.
\end{tablenotes}
\label{FRBdata1}
\end{threeparttable}
\end{table}

\clearpage

\begin{table}
\centering
\begin{threeparttable}
\caption{\textsl{Fermi}-LAT and \textsl{Swift}/BAT fluences for GRBs detected by both instrument.}
\begin{tabular}{lccccc}
\hline
\hline
Name & $T+T_{0}$\tablenotemark{*}  & Fluence BAT\tablenotemark{**}($\times 10^{-7}$)     &   Fluence LAT($\times 10^{-7}$)  \\
     & s & erg cm$^{-2}$  &  erg cm$^{-2}$ \\
\hline
GRB100728A & [6,750] & 380  & 49.6 $\pm$ 40.4 \\
GRB110625A & [0,1000] & 280 $\pm$ 10 & 133.0 $\pm$ 31.2 \\
GRB110709A & [6,42] & 100 $\pm$ 2 & 6.2 $\pm$ 3.1 \\
GRB110731A & [3,24] & 60 $\pm$ 1 & 45.8 $\pm$ 10.3 \\
GRB120624B & [100,1300] & 283 $\pm$ 4 & 127.2 $\pm$ 19.9 \\
GRB120729A & [400,800] & 24 $\pm$ 1 & 17.7 $\pm$ 8.4 \\
GRB130427A & [0,10000] & 3100 $\pm$ 30 & 3070.0 $\pm$ 298.0 \\
GRB130907A & [3000,20000] & 1400 $\pm$ 10 & 163.0 $\pm$ 116.5 \\
GRB140102A & [0,1000] & 77 $\pm$ 2 & 64.0 $\pm$ 41.6 \\
GRB140323A & [0,1000] & 160  & 25.0 $\pm$ 12.9 \\
GRB150314A & [0,250] & 220 $\pm$ 3 & 13.8 $\pm$ 5.6 \\
GRB150403A & [0,2000] & 170 $\pm$ 3 & 47.0 $\pm$ 32.0 \\
GRB150513A & [0,500] & 54 $\pm$ 2 & 26.8 $\pm$ 18.4 \\
GRB160325A & [0,2000] & 71 $\pm$ 2 & 45.6 $\pm$ 13.3 \\
GRB160821A & [0,175] & 72 $\pm$ 2 & 57.2 $\pm$ 11.3 \\
GRB160905A & [0,100] & 150 $\pm$ 2 & 36.4 $\pm$ 22.6 \\
\hline
FRB131104 & [5417,11142]  & 40$\pm$ 18   & $\le$ 19.8 \\
\hline
\end{tabular}

\begin{tablenotes}
\item\textbf{Notes.} The first column is the GRB names. The second column is the \textsl{Fermi}-LAT analysis time intervals. The third column represents the  fluences of the prompt emission detected by \textsl{Swift}/BAT. The fourth column represents  the fluences in 0.1--100~GeV detected by \textsl{Fermi}-LAT.
\item[*] $T_{0}$ is referred to each GRB trigger time.
\item[**] \url{http://Swift.gsfc.nasa.gov/archive/grb_table/}
\end{tablenotes}
\label{FRBdata2}
\end{threeparttable}
\end{table}

\end{document}